\documentclass[twocolumn,aps,pra,showpacs]{revtex4}\usepackage{epsfig}
\usepackage{graphics}
\usepackage{soul}
\begin{document}

\title{Nonclassical light from a large number of independent single-photon emitters}

\author{Luk\' a\v s Lachman, Luk\' a\v s Slodi\v cka and Radim Filip}

\affiliation{Department of Optics, Faculty of Science, Palack\' y University,\\
17. listopadu 1192/12,  771~46 Olomouc, \\ Czech Republic}

\begin{abstract}
Nonclassical quantum effects gradually reach domains of physics of large systems
previously considered as purely classical. We derive a hierarchy of operational criteria suitable for a reliable detection of nonclassicality of light from an arbitrarily large ensemble of independent single-photon emitters. We show, that such large ensemble can always emit nonclassical light without any phase reference and under realistic experimental conditions including incoherent background noise. The nonclassical light from the large ensemble of the emitters can be witnessed much better than light coming from a single or a few emitters.
\end{abstract}

\flushbottom
\maketitle

\thispagestyle{empty}

\section*{Introduction}

A stark contrast between classical and quantum physics
continuously attracts high attention of both theorists and
experimentalists. The strict and large difference in the
theoretical concepts of classical and quantum physics suggests
that no smooth transition can exist \cite{Peres}. On the other
hand, continuous and fuzzy transitions between quantum and
classical effects in many experiments provoke ongoing discussions
about an existence of quantum effects in the domain of classical
physics. The most publicly known example is the Schr\" odinger cat
experiment \cite{cat,nobel1,nobel2}, which has stimulated many
challenging experimental tests of quantum effects with a large
number of particles. Also, non-classical features of light were tested on very strong squeezed states \cite{masa}. Further, in the series of experiments in trapped ions
\cite{Monroe, Hempel}, cavity-QED experiments
\cite{Brune,Deleglise}, at optical frequencies
\cite{Bellini,Laurat}, microwave frequencies \cite{Vlastakis} and
atomic ensembles \cite{Vuletic}, the quantum correlation between a
two-level system and an oscillator has allowed to prepare a
nonclassical quantum state of a larger number of particles. The
non-classicality means here an incompatibility with oscillations
possible in classical coherence theory \cite{Glauber}. There have been developed many criteria of non-classicality, for example \cite{paris}, \cite{Vogel} etc.  However, as
the number of particles increases, many already demonstrated
nonclassical effects are critically sensitive to a coupling with
an environment \cite{Zurek}. It seems to be therefore very
challenging to observe the nonclassical effects at a truly
macroscopic limit of large ensembles, which correspond to a
traditional domain of classical physics.

Quantum optics may contribute significantly to this challenge.
Essentially, any macroscopic state of light is generated by a
large ensemble of emitters with discrete energy levels producing
individual photons. The most basic are single photon emitters,
developed recently with very small multi-photon contributions, as
was demonstrated for two-level atoms \cite{atom}, trapped ions
\cite{ion}, quantum dots \cite{dots}, NV centers \cite{nv} or
molecules \cite{molec}. Single photon states of light are
advantageously insensitive to fragile phase and provably exhibit
very robust nonclassical and even quantum non-Gaussian features
\cite{filip,jezek,lach,straka}. They remain detectable for large
optical losses, therefore suitable for a wide range of
applications. Moreover, that robust non-classicality and quantum
non-Gaussianity of light from a single emitter can be detected by
a single-photon version of the Hanbury-Brown-Twiss (HBT)
measurement \cite{mandel,grangier}. Recently, an extension of this
measurement has been used to derive a hierarchy of operational and
reliable non-classicality criteria \cite{filip1}.

In this paper we predict that arbitrary large ensemble of
independent, incoherent and even low-efficiency single photon
emitters can produce experimentally provable nonclassical light
without any phase reference and under realistic conditions. To
prove it, we modify the hierarchy of non-classicality criteria
\cite{filip1} for light from many emitters. The modified criteria
allow us to prove that if single-photon emitters individually
produce nonclassical light testable by the criteria, then source
consisting of an arbitrarily large number of these single photon
emitters generates also nonclassical light.
%Moreover, for the ensemble of low-efficient but high-quality
%single-photon emitters, the sufficient requirement for
%nonclassicality of emitted light field is the sub-Poissonian
%statistics of the number of emitters.
The proposed observation of nonclassical light is very robust
against many imperfections of the source. We verify two positive
aspects of emission from ensemble of single photon sources
compared to emission from a single one. First, the
non-classicality can be witnessed more reliably. Second,
non-classicality of a larger number of emitters is more robust
against the background noise. We suggest a feasible experimental
test of non-classicality of light emitted by large number of
trapped ions capable of verifying our theoretical predictions. %We
%propose a measurement capable to detect a small nonclassicality of
%a strong light field with photon flux
%%up to $\langle n \rangle \approx 4$
% from more than $10^4$ emitters.
%The proposed experimental test proves feasibility of measurement of
%nonclassical states of light from large number of
%single-photon emitters and allows to experimentally investigate
%fragile quantum-to-classical transition.

\begin{figure}
\centerline{\includegraphics[width=.8\linewidth]{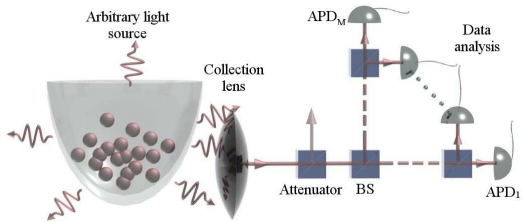}}
\caption{The proposed scheme for detection of nonclassicality.
The light emitted from many emitters is partially collimated on a
network of $50:50$ beam-splitters (BS). The non-classicality can
be detected by $M$ avalanche photodiodes (APD), each in one of the
output modes of the network. Probabilities $P_0$ and $ P_{0 \otimes M}$ are determined from the registered detection events.}
\label{figThres}
\end{figure}

\section*{Non-classicality criteria for many emitters}

Nonclassical states of light are from the point of view of
coherence theory defined as quantum states beyond the mixtures of
coherent states~\cite{Glauber}. A technique for their detection
and corresponding applied criteria have to be able to reliably and
unambiguously detect weakly nonclassical states of light without
phase reference and without any prior assumption about their
source. The detectors must be free of any systematic errors which
could cause an overestimation of the non-classicality. Moreover,
the criteria used for the tests of non-classicality have to be
applicable also for a large number of emitters.

The methodology of such operational non-classicality criteria
constructed purely from a prior knowledge about the detection
scheme presented in the Ref.~\cite{filip1} can be modified to our
purpose. The criteria can be applied to an extension of
Hanbury-Brown-Twiss type of the experiment with single photon
detectors depicted in Fig.~1. The emitted light from the source
impacts on a network of beam splitters (BS) and emerging beams are
measured by avalanche photodiodes (APDs). They can distinguish
typically with a low efficiency only photons from the vacuum. Both
the inefficiency and dark counts of the APDs reduce the
nonclassical aspects, but they cannot produce them. Ignoring them
in the construction of the criteria will not therefore cause
overestimation of the non-classicality. On the other hand, the
detection technique needs to be free of any saturation effects
which can artificially increase the estimated non-classicality.
For the same reason, it is important to measure the effective
splitting ratios including the transmissivity and reflectivity of
the beam splitters together with the generally different
efficiencies of the APDs and consider it for the derivation of the
criterion.

The non-classicality can be proven from the hierarchy of linear
functionals $P_0+aP_{0^{\otimes M}}$ of the density matrix, in
which probability $P_0$ corresponds to events when a single chosen
APD does not click and $P_{0^{\otimes M}}$ is the probability that
no click is registered by all $M$ APDs. Advantageously, we use a
parametrization by $P_0$ and $P_{0^{\otimes M}}$, different to the
one used in Ref.~\cite{filip1}, to be able to simply prove our
results and reach the best recognition of nonclassical light. We
consider now the network of beam-splitters splitting light
symmetrically to all $M$ detectors. Note that the proposed
criteria can be modified for any unbalanced network, similarly to
a simple example in Ref. \cite{filip1}. Optimizing the functional
$P_0+aP_{0^{\otimes M}}$ over all classical coherent states leads
to a threshold function
$F_M(a)=a(1-M)\left(-aM\right)^{-M/(M-1)}$, where $a \in (-\infty,
-1/M)$. This limitation arises from an optimal mean number
$\langle n\rangle_{cl}=\log\left[-aM\right]/(M-1)$ of photons in
the classical coherent state.  The linearity of the employed
functional guarantees that any convex combination of coherent
states cannot surpass the threshold $F_M(a)$ for any parameter
$a$. In turn, the existence of $a$ for which $P_0+aP_{0^{\otimes
M}}>F_M(a)$ guarantees the non-classicality of a given state. The
optimal choice of the parameter $a$ for a particular probability
$P_{0^{\otimes M}}$ corresponds to minimizing right side of
inequation $P_0>F_M(a)-a P_{0^{\otimes M}}$ over $a$. It gives
raise to an explicit form of sufficient conditions of
non-classicality
\begin{equation}
P_0^M>P_{0^{\otimes M}},
\label{nonCl}
\end{equation}
which cannot be fulfilled by any classical state of light. The
coherent states reach $P_0=P_{0^{\otimes M}}^{1/M}$. The condition
(\ref{nonCl}) for $M=2$ approximates the requirement yielded from
commonly used measurement of intensity correlation function
$g^{(2)}(0)$ for states with very low photon flux $\langle n
\rangle \ll 1$ ~\cite{mandel}. The derivation of the criterion
(\ref{nonCl}) is exact, without any approximative steps used for
example in Refs.~\cite{mandel, grangier}. It is therefore valid
even for bright sources of light, where the photon flux can not be
substituted by a probability of a click. The proposed criteria
(\ref{nonCl}) are based only on the knowledge about the detection
method, for example the effective beam splitting ratio. It is
therefore suitable to reliably detect very small non-classicality
of bright light. Importantly, inequation~(\ref{nonCl}) cannot be
satisfied by any multi-mode classical state
\begin{equation}
\rho=\int P(\alpha_1,...\alpha_i) \vert \alpha_1 \rangle \langle \alpha_1 \vert \otimes ... \otimes \vert \alpha_1 \rangle \langle \alpha_i \vert \mathrm{d}^2 \alpha_i ... \mathrm{d}^2 \alpha_i,
\end{equation}
which is a consequence of the fact, that the threshold function
$F_M(a)$ resulting from the optimization does not increase due to
the multi-mode structure. The non-classicality of light from any
source can be thus examined even for an arbitrary large number of
modes.

Suitable form of the hierarchy of non-classicality criteria
(\ref{nonCl}) allows to straightforwardly prove following result.
Consider an ensemble of independent emitters. Each emitter
gives raise to probability of no detection events $P_{0,i}$ and
$P_{0^{\otimes M},i}$, where subscript $i$ denotes probabilities
for an emitter $i$. Because emitters are considered independent,
the vacuum probability of the whole ensemble yields $P_{0^{\otimes
M}}=\Pi_{i=1}^N P_{0^{\otimes M},i}$ and $P_{0}=\Pi_{i=1}^N
P_{0,i}$. There is a connection between non-classicality of a
single emitter and the non-classicality of whole ensemble
following from the statement
\begin{eqnarray}\label{theo}
&\ & (\forall i:\ P_{0,i}^M>P_{0^{\otimes M},i} )\Rightarrow \nonumber \\
&\ &(\Pi_{i=1}^N P_{0,i})^M >\Pi_{i=1}^N P_{0^{\otimes M},i} \Leftrightarrow P_{0}^M>P_{0^{\otimes M}}
\end{eqnarray}
for arbitrary $M \geq 2$, where $i=1,\ldots,N$. Therefore, any
ensemble of independent emitters all individually satisfying
(\ref{nonCl}) generates arbitrarily bright nonclassical light
detectable by the same criterion (\ref{nonCl}). Due to a very
suitable parametrization of the problem, the expression can be
proved by simple algebraic rules. A main task was to find the most
suitable parametrization of non-classicality criteria, so that the
proof is simple and broadly understandable. Although participating
$N$ single-photon emitters can be quite different, the
non-classicality of light emitted by each emitter suffices to
obtain a nonclassical state yielded from arbitrarily large
ensemble of them. In the Supplementary Material, we discuss a
tolerance of the nonclassical light emission from realistically
unstable large ensembles.

\section*{Reliability of nonclassical criteria}

To obtain a conclusive proof of non-classicality, the statistical
errors of the measured distance from the non-classicality
threshold have to be sufficiently small. The necessity of small
errors dictates in turn a minimal time needed for such
experimental estimation. We have chosen a logarithmic scale
for its better interpretations. The criteria (\ref{nonCl}) get a
linear condition $M \log_{10} P_0 < \log_{10} P_{0^{\otimes M}}$.
We consider a simplest example of an ensemble of $N$ single photon
emitters, where any emitter produces maximally a single photon in
the state $\rho_{S}=\eta|1\rangle\langle
1|+(1-\eta)|0\rangle\langle 0|$. The probabilities for the
criterion (\ref{nonCl}) are $P_0=\left(1-\eta/M\right)^N$ and
$P_{0^{\otimes M}}=\left(1-\eta\right)^N$. Let us define distances
along the axes for states above the threshold
\begin{eqnarray}
d_0=\log_{10} P_0-\log_{10} \sqrt[M]{P_{0^{\otimes M}}}  \nonumber \\
d_{0^{\otimes M}}=\log_{10} P_0^M-\log_{10} P_{0^{\otimes M}}.
\label{distG}
\end{eqnarray}
The convenient choice of the log-log space for the
description of the distance from the threshold (\ref{distG})
connects both distances $d_{0^{\otimes M}}$ and $d_0$ so that
$d_{0^{\otimes M}}=M d_0$. From the point of view of measurement,
the relevant distance $d$ from the threshold is given by
$d=\sqrt{d_{0^{\otimes M}}^2+d_0^2}$, which implies
$d=\sqrt{M^2+1}d_0$. If the state does not satisfy the condition
of non-classicality (\ref{nonCl}), we assign zero to both
distances $d_0=d_{0^{\otimes M}}=0$ to avoid obtaining negative
distance. We stress, that this distance does not quantify an
amount of non-classicality.  The distance for $N$ emitters
producing individually $\rho_S$ with the same efficiency of
generation $\eta$ is
\begin{eqnarray}
d & =&N \sqrt{M^2+1} \log_{10}\frac{\left(1-\frac{\eta}{M}\right)}{\sqrt[M]{1-\eta} }\nonumber \\
&\approx & N\frac{(M-1)\sqrt{M^2+1}}{2M^2\ln 10}\eta^2
\label{dist}
\end{eqnarray}
where the approximation is valid for typical low-efficient
emitters with $\eta \ll 1$. {\em Interestingly, light from many
emitters increases the distance of the point from the
non-classicality threshold}. As can be seen in Fig ~2, the
distance is very sensitive to the single photon emitter efficiency
$\eta$. Furthermore, the presented plot suggests, that small
number of the experimental runs will suffice to obtain reliable
estimate about non-classicality of emitted light. For larger
number of detectors $M$, the distance $d$ saturates to $d\approx
N\frac{\eta^2}{2\ln 10}$ for small $\eta\ll 1$. The small
$\eta^2$ can be advantageously compensated by a large number of
emitters $N$. This opens possibility to detect non-classicality
using only a fraction of emitted light.
%We have to therefore optimize $M$ not to be very close from the threshold along the axis with $d_0$.

\begin{figure}
\centerline{\includegraphics[width=.8\linewidth]{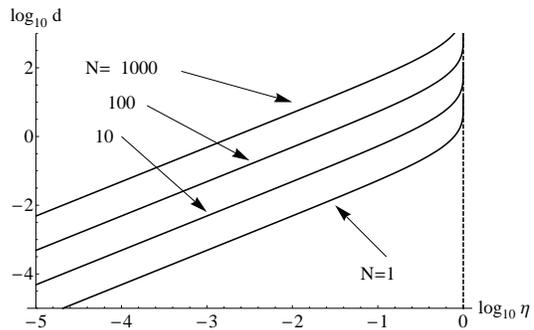}}
\caption{The reliability of the simplest criterion with $M=2$
detectors in (\ref{nonCl}) is depicted for the discussed model
$\left[\eta \vert 1 \rangle \langle 1 \vert + (1-\eta) \vert 0
\rangle \langle 0 \vert \right]^{\otimes N}$. The reliability is
qualified by the proposed distance $d$ of a point $(P_0,\
P_{0^{\otimes 2}})$ from the threshold using equation
(\ref{dist}). It affirms the advantage of using a large ensemble
of single photon states.} \label{figmacro}
\end{figure}

\section*{Background noise}

Since $P_{0}^M=P_{0^{\otimes M}}$ for the classical noise with
Poissonian statistic, any source of multi-mode Poissonian noise
added to the emitters of nonclassical light keeps non-classicality
observable by (\ref{nonCl}), what can be simply proven using
(\ref{theo}). On the other hand, thermal incoherent background
noise limits detection of nonclassical light~\cite{lach}. The
influence of the thermal noise on light from $N$ single photon
emitters can be described by two principal models
 \begin{equation}
 \rho^{(N)}_{1}=\left[\rho_{S} \otimes \rho_{TH}(\bar{n})\right]^{\otimes N},\,\,\,\rho^{(N)}_{2}=\rho_{S}^{\otimes N} \otimes \rho_{TH}(\bar{n}),
 \label{noise}
 \end{equation}
where
$\rho_{TH}(\bar{n})=\sum_{n=0}^{\infty}\bar{n}^{n}/(1+\bar{n})^{n+1}|n\rangle\langle
n|$ is thermal noise with a mean photon number of $\bar{n}$. We
investigated all criteria (\ref{nonCl}) and numerically verified
that {\em the basic criterion for $M=2$ is sufficient} for both
$\rho^{(N)}_{1}$ and $\rho^{(N)}_{2}$ for arbitrary large $N$. It
demonstrates that detection setup for nonclassical light from the
large ensemble can be very simple. In many cases, we can use the
simplest case with $M=2$, therefore we will further focus on this
simplest case.

In the state $\rho_1^{(N)}$, which corresponds to the noise
contributing to each single emitter, non-classicality appears if
$\eta>\bar{n}/(1+\bar{n})$. This condition holds for arbitrary
large number of emitters. On the other hand, for the right state
$\rho_2^{(N)}$, which describes the background noise jointly
affecting all emitters, the threshold for non-classicality
interestingly decreases for a large number of emitters $N$. For
very low noise $\bar{n}\ll 1$, the nonclassical light is detected
if
 \begin{equation}
 \eta>\frac{\bar{n}}{\sqrt{N}}.
 \label{BEapp}
 \end{equation}
It is a clear application of many emitters, they can be used to
{\em detect non-classicality not measurable from a single
emitter}. For the state $\rho_1^{(N)}$, non-classicality tolerates
optical attenuation when $T>\frac{\bar{n}-\eta}{\eta \bar{n}}$,
irrespectively to $N$. For $\eta>\bar{n}$, the non-classicality
cannot be destroyed by attenuation and it is absolutely robust.
%If only the number of the single photon emitters decay, the condition changes to more strict $\eta>\bar{n}e^{t/\sqrt{\tau}}$.
%On the other hand, if the number of emitters decays simultaneously with the sources of noise, the condition $\eta>\bar{n}$ remains.
For the state in $\rho_2^{(N)}$, the approximate condition (\ref{BEapp}) appeared to be also for the absolute robustness.
%Non-classicality conditions gained here for a fixed number of single photon emitters $M$ can be easily applied on a decaying number of emitters, using $\eta \rightarrow \eta e^{-t/\tau}$, together with either the sources thermal background unchanged or decaying as well.

%Thus, the condition to observe nonclassicality is $\eta>\frac{\bar{n}}{1+\bar{n}}e^{t/\sqrt{\tau}}$
%for the sources of thermal background noise in each emitter which do not decay. The small $\bar{n}\ll 1$ is effectively increased by decaying number of the particles. It can therefore limit the detectability of nonclasssical light from many emitters with background noise in each emitter.
%If the noise sources decay simultaneously with the emitters, the condition changes to
%$\eta>\frac{\bar{n}}{1+\bar{n}e^{-t/\sqrt{\tau}}}$,
%because the initial $\bar{n}$ reduces by the same exponential decay. It simply converges to condition $\eta>\bar{n}$ for any $\bar{n}$. For the joint thermal noise, the aprroximate nonclassicality condition $\eta>\frac{\bar{n}}{\sqrt{M}}e^{t/\sqrt{\tau}}$.
%says that the decay is not as limiting for the large number of emitters.

\begin{figure}
\centerline{\includegraphics[width=\linewidth]{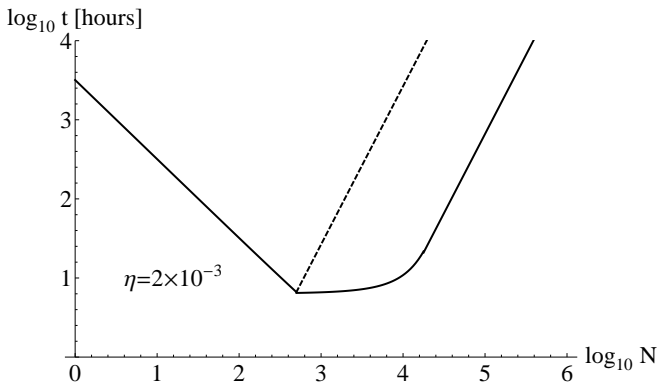}}
\caption{Estimation of the minimal time needed for the
measurement of non-classicality of light from $N$ trapped ions
corresponding to reaching the distance $d$ more than three
standard deviations from the non-classicality threshold. The solid
lines show simulation results employing attenuations of the
measured signal using both periodic switching-off of the detectors
and beam splitter attenuation, while the dashed lines corresponds
to the case where detectors saturation is avoided only by more
conventional beam-splitter method.} \label{simulationX}
\end{figure}

\section*{Measurement time}

A remaining limit for the detection of nonclassical light seems to
be only finite measurement time.
The distance from the threshold (\ref{nonCl}) for $M=2$  is
depicted in the Fig.~(\ref{figmacro}). The distances have to be
smaller than demanded number of standard deviation units obtained
by
\begin{equation}
\delta \log_{10} P_{0^{\otimes 2}}\propto \sqrt{\frac{1-P_{0^{\otimes 2}}}{P_{0^{\otimes 2}} t_m}}\approx \frac{e^{N \eta/2}}{\sqrt{t_m}},
\label{sigma}
\end{equation}
where $t_m$ is measurement time. The approximation is correct only with assumption $\eta \ll 1$
and $P_{0^{\otimes 2}} \ll 1$. For a large number of emitters and
corresponding large photon flux, the measurement time can be
optimized by attenuation of photon flux. The attenuation with
efficiency $T$ modifies the probabilities to $P_{0}=1-T\eta/2$ and
$P_{0^{\otimes 2}}=1-T\eta$. Clearly, any fixed number of these
sources always radiates provably nonclassical light preserved
through any attenuating channel. The attenuation can be optimized
to reach sufficiently high $P_{0^{\otimes 2}}$ and simultaneously,
not too small distance from the threshold. According to
(\ref{dist}) and (\ref{sigma}), it appropriates to optimization of
the measured time $t_m \propto N^2\exp(T N \eta)/(T N \eta)^4$.
One can find an optimal photon flux $T N \eta$ and the
corresponding increase of time is quadratic in number of emitters.
On the contrary, the necessary measurement time grows
exponentially without performing the optimal attenuation. Since an
arbitrarily large ensemble of such emitters generates
non-classical light, its detection reduces to search of relevant
experimental scheme and parameters optimization.

\section*{Experimental proposal}

A large number of trapped atoms or ions \cite{Drewsen, Schifffer}
can be advantageously used for an experimental demonstration. We
consider a system containing $N$ ions in a cigar-like shaped
crystal \cite{Werth} trapped in a focus of a lens with a numerical
aperture of 0.2 corresponding to 1\% of the full solid angle being
continuously driven by excitation laser. We estimate a minimal
measurement time needed to proof non-classicality of emitted
light, which we set to time needed for reaching $d/\sigma_d=3$,
where $\sigma_d$ is one standard deviation of measured distance
$d$. We assume that the number of ions in the trap is constant,
which is justified if $t_m \ll \tau_s/\sqrt{N}$, where $t_m$ is
measurement time, $\tau_s$ is storage time of an ion in the trap
and $N$ is a number of ions. Thus for very conservative storage
time $\tau_s=1$~day and an ensemble containing several tens of
ions, the measurement should no exceed one hour. If the required
measurement time exceeds this assumption, the experiment can to be
repeated several times with shorter time duration and with same
initial number of emitters, so that statistics of decaying number
of emitters is included in the measured vacuum probability.
Collected photons are collimated and directed towards two APDs
through variable attenuator and an ordinary 50:50 BS as depicted
in Fig ~1. The emitted fluorescence collected by the lens %In figure \ref{experiment}-a) is shown the dependence
%of surpassing the non-classicality threshold by three standard
%deviations as a function of the measurement time for different
%average dark counts of employed detectors for a
is then detected with very modest overall detection efficiency
$\eta$=0.2~\%, which corresponds to realistic value of single
photon detector quantum efficiency of 50~\%, 50:50 splitting ratio
of the beam splitter and additional 20~\% absorbtion and
reflection losses at optical elements. Considering the excitation
and detection of the photons from the dipole transition of some
frequently used alkaline earth metal ions with the excited state
lifetime $t_{e}\sim 100 {\rm ns}$, we set the length of the single
measurement time-bin to $t_b=10 {\rm ns}\ll t_e$ to minimize the
multi-photon contributions from the same ion in the single
time-bin. The efficiency $\eta$ can be further artificially
decreased by the variable attenuation with factor $T$. This
attenuation strongly influences the measurement duration and one
can find optimal $T<1$ which minimizes the measurement time. The
estimated optimal photon flux per single time bin in our case
corresponds to $N T \eta \approx 4.1$. The probability of a click
for such flux is $1-P_{0^{\otimes 2}} \approx 0.98$. As can be
seen in the simulation results presented in
Fig.~\ref{simulationX}, our approach enables to demonstrate
non-classicality of emitted light within a $100$ hours for an
ensemble of up to $N \sim 10^4$ ions, which approaches a
truly macroscopic limit.

In the presented simulation  we account for the effect of
saturation of conventional single photon detectors which becomes
substantial typically at a count-rate of about $500\,$kHz, what
corresponds approximately to $\eta N \sim 0.05$. For the photon
flux of the source below this value, the APDs can be opened per each temporal
mode without being saturated. As can be seen in the
Fig.~\ref{simulationX}, the minimal measurement time in this
region rapidly decreases due to the simultaneous increase of the
distance $d$ and measured photon flux as the number of emitters
$N$ grows. Above this region, we start avoiding the saturation of
detectors by keeping both APDs switched-off most of the time and
opening them only for~$10\,{\rm ns}\sim t_b$ periods with a
frequency of the opening given by the saturation limit. Activation
of this mechanism is responsible for a strong kink in the
dependence of the minimal measurement time on the number of
particles at $\eta N \sim 0.05$. 
Further increase of the number of
particles is compensated by decreasing the overall time in which
APDs are switched on. The minimal measurement time in this region
stays approximately constant up to the point, where the frequency
of APD switching-off reaches 500~kHz, which corresponds to regime
where one photon is detected at each APD in almost each detection
period. We note, that we don't consider a detection of more than
one photon by single APD in a given detection interval, which is
prohibited by dead time $\tau_{\rm DT}$ of APDs being typically
$\tau_{\rm DT}\gg t_b = 10$\,ns. This assures, that further
increase of number of emitters will not cause saturation in this
regime. When compared to conventional beam-splitter type of
attenuation to obtain the regime below detector's saturation which
is depicted by a dashed line in Fig.~\ref{simulationX}, this
approach substantially reduces total measurement time. This is
mainly caused by the fact, that an estimated optimal value of the
$\eta N \approx 4.1$ is far behind the saturation limit of
conventional single photon detectors. Further increase of the
number of emitters and corresponding photon flux behind this value
is then compensated by usual beam-splitter type of attenuation
with attenuation factor $T$ which gives a quadratic increase of
measurement time.

\section*{Application and outlook}

Practically, our methodology can be generally applied as an efficient tool enabling unambiguous searching for
nonclassical behavior hidden in various experimental platforms without
any prior knowledge. A potential of nonclassical states of large ensemble can be further found in quantum metrology and quantum communication. The proposed experiment will likely move forward ongoing investigation of quantum to classical transitions with large quantum or even mesoscopic systems.

\section*{Methods}

%%%%%%%%%%%%%%%%%%%%%%%%%%%%%%%%%%%%%%%%%%%%%%%%%%%%%%%%%%%%
\subsection*{Statistics of emitters}

Statistics of the number of emitters clearly influences the
non-classicality of emitted radiation. If the emitters generate
exactly single photon, the thermal (Poissonian) statistics of
emitters will produce thermal (Poissonian) statistics of light. On
the other hand, the emitters producing states
$\rho=\eta|1\rangle\langle 1|+(1-\eta)|0\rangle\langle 0|$ can be
measured as nonclassical using the hierarchy of criteria (Eq.~(1)
in the main text) only if
\begin{equation}\label{cond}
\langle (1-\frac{\eta}{M})^N\rangle^M>\langle (1-\eta)^N\rangle,
\end{equation}
where the averaging is over the fluctuating number of emitters.
Inserting low efficiency approximation $\langle (1-\eta)^N\rangle\approx 1-\eta\langle N\rangle+\frac{\eta^2}{2}\left(\langle N^2\rangle-\langle N\rangle\right)$ and $\langle (1-\frac{\eta}{M})^N\rangle\approx 1-\frac{1}{M}\eta\langle N\rangle+\frac{\eta^2}{2M^2} (\langle N^2\rangle-\langle N\rangle)$ with $\eta \ll 1$ into the hierarchy of criteria (Eq.~(1) in the main text) and considering terms only up to $\eta^2$, we obtain that the variance
\begin{equation}\label{subPois}
V(N)=\langle N^2\rangle - \langle N\rangle^2<\langle N\rangle
\end{equation}
has to be smaller than Poissonian to obtain the nonclassical light
detectable for any $M \geq 2$.

The approximation of small $\eta\ll
1$ for (\ref{subPois}) has been verified numerically and compared
to (\ref{cond}). Sub-Poissonian statistics of single photon
emitters therefore suffices to generate detectable
nonclassical light. For the simplest scheme with $M=2$ and the ensemble of imperfect single-photon
emitters producing an approximative state $\rho\approx
(1-\eta_1-\eta_2)|0\rangle\langle 0|+\eta_1|1\rangle\langle
1|+\eta_2|2\rangle\langle 2|$, where $\eta_1,\eta_2\ll 1$ are
small single-photon and two-photon contributions, we can
approximatively obtain the condition
$V(N)<\left(1-\frac{2\eta_2}{\eta_1^2}\right)\langle N\rangle$.
The ratio $\frac{2\eta_2}{\eta_1^2}$ corresponds to the ratio
$\frac{P_C}{P_S^2}$ converging to the correlation function
$g^{2}(0)$ for $\eta_1,\eta_2\ll 1$, where $P_C\approx \eta_2/2$
is a probability of a coincidence detection
and $P_S\approx \eta_1/2$ is approximately a probability of a
detection event at the single detector. Non-vanishing
$g^{2}(0)$ therefore suppresses the upper bound on $V(N)$ below
the Poissonian limit. If the number of emitters is
controlled reasonably below Poissonian limit, the detection of nonclassical
states from a large number of emitters is feasible.

%%%%%%%%%%%%%%%%%%%%%%%%%%%%%%%%%%%%%%%%%%%%%%%%%%%%%%
\subsection*{Decaying number of emitters}

For the test of non-classicality of emitted light from $N$
emitters, it is not needed to know the number of emitters. It is
ideal for the test, however, the number of single photon emitters
exponentially decays with relatively small time constant in many
experimental platforms. They can either gradually leave the
ensemble to the environment or they can gradually loose
possibility to emit light. Let us consider an ensemble that
initially contains $N$ single photon emitters radiating maximally
a single photon with probability $\eta$. Further, we assume that a
probability that an emitter radiates in the ensemble at time $t$
is $P(t)=\exp(-t/\tau_s)$, where $\tau_s$ is storage time of the
emitter in the ensemble. The photon distribution of such source in
time is
\begin{equation}
P_n(t)=\sum_{k=n}^N{ N \choose k}{ k \choose n} e^{-k t/\tau_s}(1-e^{-t/\tau_s})^{N-k}\eta^n(1-\eta)^{k-n}.
\label{fluctProb}
\end{equation}
According to (\ref{fluctProb}), it is possible to understand the
losses of emitters from the ensemble as an attenuation of emitted
light with the transmittancy $T=e^{-t/\tau_s}$. Thus, the
fluctuating number of emitters can be included in the
efficiency of the individual emitters $\eta' = \eta e^{-t/\tau_s}$,
which enters the vacuum probabilities
$P_{0^{\otimes M}}=\left(1-\eta'\right)^N$ and
$P_{0}=\left(1-\eta'/M\right)^N$. Since $\eta\rightarrow\eta'$,
all previous conclusions for $\eta>0$ remain valid for $\eta'>0$,
if the measurement starts in time $t_0$ and lasts for time $t_m$
much shorter than  $t_m \ll \tau_s/\sqrt{N}$ hence the number of emitters is
stable during measurement. However, long measurement time or a large number of
emitters violate this assumption and the detected vacuum
probability has to be averaged
\begin{equation}
\langle P_{0^{\otimes M}} \rangle = \frac{1}{t_m}\int_{t=t_0}^{t_0+t_m} \left(1-\eta e^{-t/\tau_s}\right)^N \mathrm{d}t.
\label{avrVac}
\end{equation}
The average probability $\langle P_0 \rangle$ is yielded by
substitution $\eta \rightarrow \eta/M$ in (\ref{avrVac}). The
view, how this averaging limits the observation of
non-classicality, is reached in an approximation of very weak
sources $\eta \ll 1$, where the integration is analytically
achievable. The first two members in Taylor expansion are
$P_{0^{\otimes M}}(t)\approx 1-N \eta'(t)+N(N-1) \eta'(t)^2 /2 $
with $\eta'(t)=\eta \exp(-t/\tau_s)$. To expand $P_0(t)$, we can
substitute $\eta\rightarrow \eta/M$ in the expansion of
$P_{0^{\otimes M}}(t)$. The subtraction $\langle P_0
\rangle^M-\langle P_{0^{\otimes M}} \rangle$ cancels the parts
proportional to $\eta$ automatically. The second parts
proportional to $\eta^2$ do not contribute to the expansion only
if $N=t_m/\left[t_m-2\tau_s
\tanh\left(\frac{t_m}{2\tau_s}\right)\right]$ for any $M\geq 2$.
Consequently, it dictates condition on a maximal number of single
photon emitters such that the measurement duration $t_m$ does not
cause lost of non-classicality for $\eta \ll 1$ and any $M\geq 2$
\begin{equation}
N<\frac{t_m}{t_m-2\tau_s \tanh\left(\frac{t_m}{2 \tau_s}\right)}.
\label{Nmax}
\end{equation}
The small time $t_m$ compared to the storage time gives
approximative condition $N<\tau_s^2/t_m^2$. The lost of visibility
of non-classicality can be avoided by repeating the measurement
several times with the same initial number of emitters.

%%%%%%%%%%%%%%%%%%%%%%%%%%%%%%%%
%Since $\eta\rightarrow\eta'$, all previous conclusions for $\eta>0$ remain valid for $\eta'>0$ as well. We can therefore omit the prime symbol in $\eta'$ and consider $\eta$ includes both the efficiency of generation and the decay of a number of the emitters. The decay of definite numbers of many single-photon emitters therefore does not break emission of nonclassical light.

%%%%%%%%%%%%%%%%%%%%%%%%%%%%%%
\subsection*{Fluctuating single-photon efficiency}

The efficiency of generation and detection of emitted single
photons from individual emitters $\eta$ can fluctuate during the
measurement time, although the number of emitters is definite. The
probability of vacuum is then $P_{0}=\langle \Pi_{i=1}^N
(1-\eta_i/M) \rangle$ and $P_{0^{\otimes M}}=\langle \Pi_{i=1}^N
(1-\eta_i) \rangle$, where $\eta_i$ is the random efficiency of
generation in $i$th emitter and $\langle ...\rangle$ means mean
value. Let all the emitters have
$\langle\eta_i\rangle=\langle\eta\rangle$. Considering that any
two emitters do not influence each other, we use $\langle\eta_i
\eta_j \rangle=\langle\eta_i \rangle \langle\eta_j \rangle$,
$i,j=1,\ldots,N$ to obtain $P_0=(1-\frac{\langle \eta
\rangle}{M})^N$ and $P_{0^{\otimes M}}=(1-\langle \eta
\rangle)^N$. It clearly satisfies $P_0^M> P_{0^{\otimes M}}$. The
different and fluctuating $\eta$ of single photon emitters thus
cannot hinder the observability of nonclassical light. We can
detect non-classicality for the smallest $M=2$, however, the
distance from the threshold increases for larger $M$.

%%%%%%%%%%%%%%%%%%%%%%%
%\begin{figure}[t]
%\centerline{\includegraphics[width=0.5\textwidth]{thermal}}
%\label{thermal}
%\caption{The non-classicality of source with %background noise having Bose - Einstein %distribution of photons is depicted by solid %line. The dashed line is linear approximation of %threshold valid in the very attenuated state %regime.}
%\end{figure}
%%%%%%%%%%%%%%%%%%%%%%%%%%

\section*{Acknowledgements}

We acknowledge financial support from grant No. GA14-36681G of the Czech Science Foundation. L.L. acknowledges the support of Palack\'y University (IGA-PrF-2015-005).

\section*{Author contributions}

R.F. provided the idea, theoretical concept and general analysis. L.L. mainly contributed to detailed theoretical calculations and analysis of measurement. L.S. suggested experimental test and analysis of measurement. All authors were involved in writing, editing and revising the manuscript.

\section*{Additional Information}

\paragraph{Competing financial interests:} The authors declare no competing financial interests.
\end{document}